\RequirePackage{fix-cm}
\documentclass[twocolumn]{svjour3}          
\smartqed  
\usepackage{graphicx}
\usepackage{makecell}
\usepackage{subfigure}
\usepackage{float}
\usepackage{verbatim}
\usepackage{lineno}

\begin{document}
	

    \title{Muon reconstruction with a convolutional neural network in the JUNO detector
    			\thanks{Supported by Strategic Priority Research Program of Chinese Academy of Sciences (XDA10010900) and NSFC (11805223)}
    } 

	
	
	\author{Yan Liu \and Wei-Dong Li \and Tao Lin \and Wen-Xing Fang \and Simon C. Blyth \and Ji-Lei Xu \and Miao He \and Kun Zhang}
	
	\institute{
      Yan Liu \and Wei-Dong Li
      \at University of Chinese Academy of Sciences, Beijing 100049, China
      \and
      Yan Liu \and Wei-Dong Li \and Tao Lin \and Wen-Xing Fang \and Simon C. Blyth \and Ji-Lei Xu \and Miao He \and Kun Zhang
      \at Institute of High Energy Physics, Chinese Academy of Sciences, Beijing 100049, China \\\email{lintao@ihep.ac.cn}
	}
	
	\date{Received: date / Accepted: date}

	\maketitle
	
	\begin{abstract}
		\textbf{}\\

The Jiangmen Underground Neutrino Observatory (JUNO) is designed to determine
the neutrino mass ordering and measure neutrino oscillation parameters.  
A precise muon reconstruction is crucial to reduce one of the major backgrounds
induced by cosmic muons.
This article proposes a novel muon reconstruction
method based on convolutional neural network (CNN) models. In this method, the
track information reconstructed by the top tracker is used for network
training. The training dataset is augmented by applying a rotation to muon tracks
to compensate for the limited angular coverage of the top tracker.
The muon reconstruction with the CNN model can produce unbiased tracks with performance that spatial resolution is better than 10 cm and angular resolution is better than 0.6 degrees. 
By using a
GPU accelerated implementation a speedup factor of 100 compared to existing CPU
techniques has been demonstrated.  
	\end{abstract}
	
	\section*{Introduction}
	\label{intro}
\begin{figure}[htbp]
    \centering  
    \includegraphics[width=0.7\linewidth]{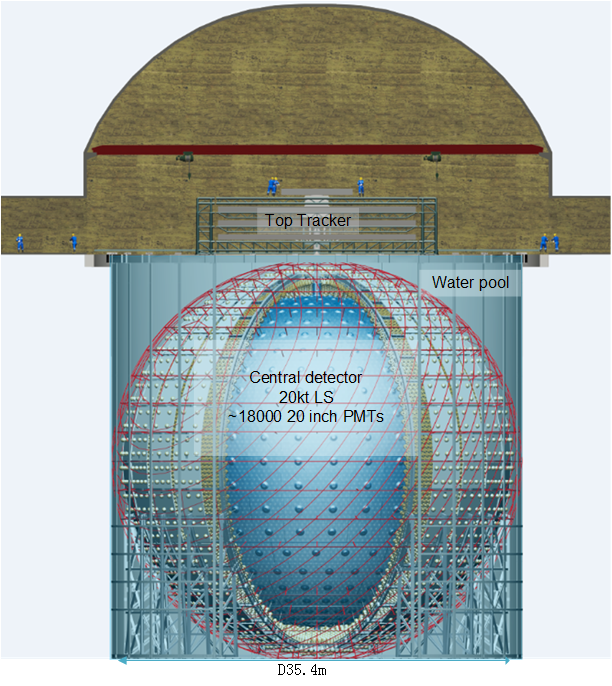}  
    \caption{A schematic view of the JUNO detector.}
    \label{fig:1}   
\end{figure}

Jiangmen Underground Neutrino Observatory (JUNO) \cite{Djurcic:2015vqa} is a 
liquid scintillator detector that is designed to determine the neutrino mass
ordering and to precisely measure neutrino oscillation parameters. Its location 53 km away 
from both Yangjiang and Taishan nuclear power plants in Southern China 
is chosen to optimize the oscillation measurement sensitivity.
A schematic view of the JUNO detector is shown in Fig.~\ref{fig:1}. The innermost
central detector (CD) is 20 kiloton of liquid scintillator contained by an acrylic sphere 
and instrumented with $\sim$18,000 20-inch and 25,600 3-inch photomultipliers tubes (PMTs).
The CD is submerged in a water pool 
which is instrumented with a further 2,400 PMTs forming the Water Cerenkov Detector (WCD).
The Top Tracker (TT), located above the CD and WCD, is a 3-layer muon tracker detector with size 47 m $\times$ 20 m $\times$ 3 m providing precise atmospheric muon tracking over about 60\% of the water pool.
Ionizing particles cause excitations within the scintillator which 
subsequently de-exite resulting in the emission of scintillation photons. 
These photons propagate through the detector undergoing scattering, absorption and re-emission
as well as refraction and reflection at material boundaries before some reach the PMTs. 
The charge and time information measured by the PMTs enables the originating 
vertex position and energy to be reconstructed.
%
%
%
%

The primary background for neutrino detection arises from $^{9}$Li/ $^{8}$He induced by
muon spallation in the liquid scintillator or muon shower particles. This
background can be reduced by excluding a veto volume along the muon
track within a time window surrounding the time at which the muon 
passes through the detector. 
A precise muon reconstruction directly allows the veto volume and time 
windows to be reduced which increases the livetime achievable by 
the JUNO detector. 

The vital importance of muon reconstruction for JUNO has led to the development 
of several muon reconstruction methods including:
\begin{itemize}
\item Muon reconstruction with a geometrical model~\cite{Genster:2018caz} which is 
based on the geometrical shape of the first light and provides
a spatial resolution of 20 cm and an angular resolution of 1.6 degree 
over the whole detector.
\item Muon tracking with the fastest light~\cite{Zhang:2018kag}
which uses the least squares fit to the PMT first hit time (FHT) measurements. 
After correction of the FHT for each PMT, the reconstruction yields
spatial resolution less than 3 cm and angular resolution of less than 0.4 degree. 
\end{itemize}

This article reports a novel alternative technique for muon reconstruction 
that applies deep learning techniques and benefits from GPU acceleration.
This technique has the advantages of speed and avoids
the need to develop detailed optical models or to interpret calibration data.
All studies have been based on JUNO Monte Carlo simulation data. 

Treating the readout of the whole detector as an image composed of pixel values from
PMT charge and time measurements allows image classification techniques to be applied to track finding and fitting. 
For image processing tasks, Convolutional Neural Networks (CNN) have been widely used in many areas such as image classification \cite{Russakovsky:2014lic}, object detection \cite{girshick2014rich} and image
segmentation \cite{long2015fully}. Within high energy physics, image processing with CNN has been used 
in jet classification \cite{Kasieczka:2017nvn,Baldi:2016fql} and event classification \cite{Bhimji:2017qvb,Aurisano:2016jvx}. In this paper, the CNN is adopted to perform muon track reconstruction in the CD detector.

This paper is organized into a description of how the training and testing datasets were prepared 
in section 2 followed by an explanation of the chosen deep neural network architecture 
and training procedure in section 3. Performance metrics of the muon reconstruction are presented in section 4 
and a discussion of possible challenges for the application of this technique to experimental 
data are given in section 5, followed by the conclusions of this study summarized in section 6.

\section*{Dataset and muon reconstruction model}%
\label{sec:1}%
\subsection*{Simulation of single muon tracks}
In this study all Monte Carlo (MC) simulated muon events used for network training and testing are produced by the official JUNO simulation framework~\cite{Huang:2017dkh,Lin:2017usg} which is based on the Geant4 toolkit~\cite{allison2016recent}. The underground muon sample before input into detector are generated by the muon transportation software MUSIC~\cite{Kudryavtsev:2008qh} with JUNO mountain profile and the parameterized Gaisser formula of muon \cite{Guan:2015vja} at the mountain surface. The energy range of cosmic muon reaching the JUNO detector is from 0.1 GeV to 10 TeV in simulation and the average energy is about 207 GeV. The flux is about 4 $\times$ 10$^{-3}$ Hz / m$^{2}$. There are about 90$\%$ muons with single track and about 10$\%$ muons with multiple tracks in the muon events. Only the events with 200 GeV muon track are selected for this study, so that the muons will go through the whole detector and exit at the bottom.

Along the muon trajectory through the materials of the detector geometry 
the Geant4 toolkit simulates physics processes including ionization and nuclear interactions.
Muons passing through the scintillator result in excitations of the molecules of the 
scintillating medium which subsequently de-excite producing photons. 
In addition as muons travel through both the water pool and the scintillator their 
charged nature results in the generation of photons from Cerenkov radiation.
Both scintillation photons and Cerenkov photons are propagated through the detector 
taking into account optical processes such as scattering, absorption and re-emission. 
A fraction of the propagated photons reach PMTs and eject a photo-electron
from the photocathode resulting in electronic signals for the charge and arrival time.
PMT characteristics including the quantum efficiency and collection efficiency are measured and used as inputs to 
the simulation.
	
\begin{figure}[h]
    \centering  
    \includegraphics[width=0.3\textwidth]{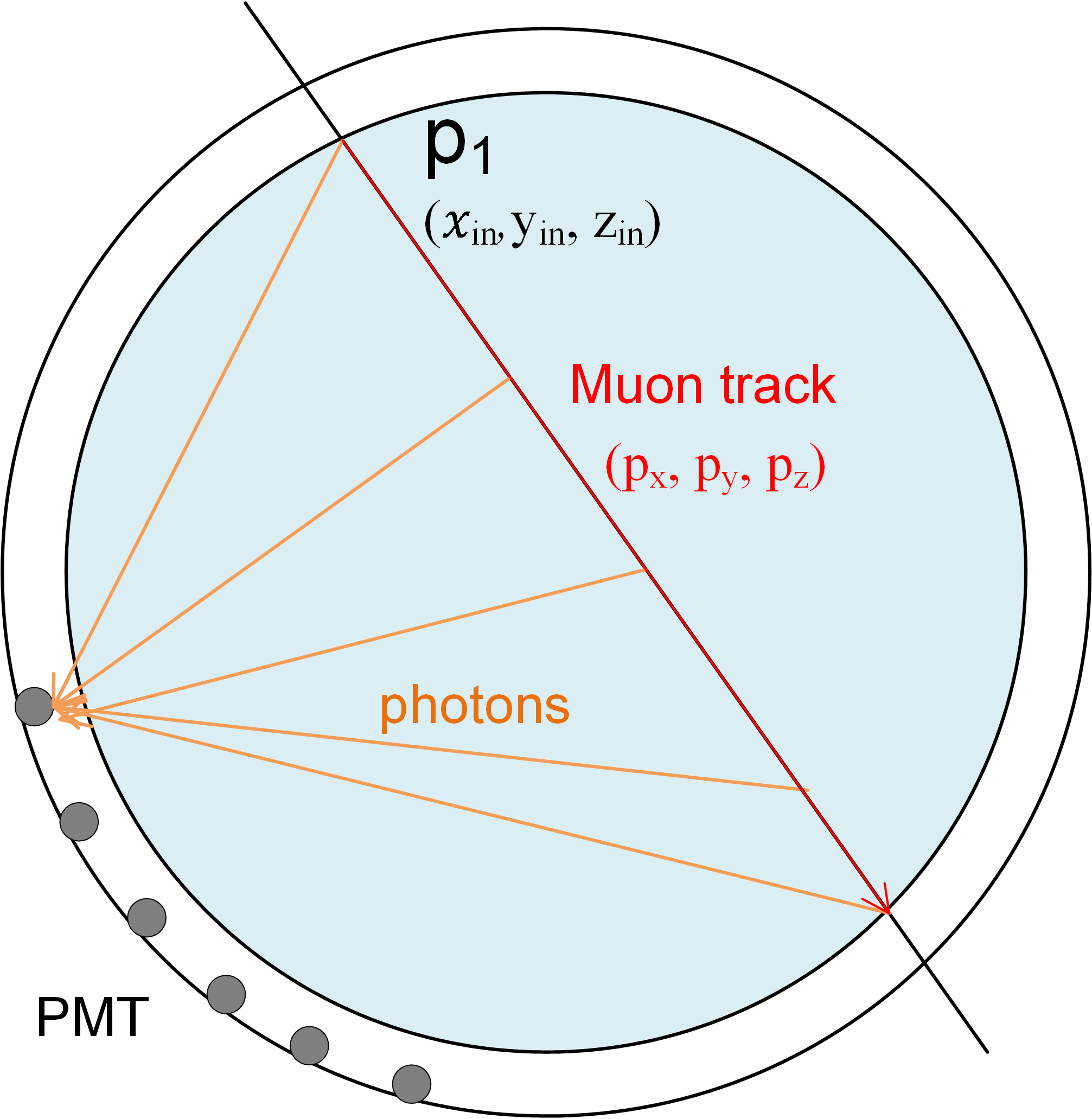}
    \caption{Schematic of muon track, optical photon paths and PMT positions. The shaded area represents the LS.} 
    \label{fig:2}
\end{figure}

Figure~\ref{fig:2} illustrates a muon track passing through the liquid scintillator (LS),
together with the positions of PMTs that detect optical photons emitted along the muon
trajectory. 
	
Comparisons of simulated and real experimental data often reveal some differences.
This presents a problem as neural networks trained with simulated samples 
may learn features that differ from those present in the real data 
causing the model to be invalid. 
As real JUNO experimental data is not yet available a comparison 
between independent aspects of the simulated data is used as a cross check.
The muon track parameters obtained from the top tracker (TT) reconstruction~\cite{Djurcic:2015vqa} 
are used as an independent alternative for training the neural network and compared
with results obtained by training based on the MC truth information. 

\subsection*{Modeling of muon reconstruction}

Muon reconstruction uses the first light arrival times and number of 
photoelectrons detected by all the PMTs as well as the positions of the PMTs.
The muon track is modeled using its direction vector ($p_{x},p_{y},p_{z}$) 
and the intersection position of the track with surface of the LS sphere 
$\vec{r}_{\scriptsize\textrm{in}}$ = ($x_{\scriptsize\textrm{in}},y_{\scriptsize\textrm{in}},z_{\scriptsize\textrm{in}}$) using a coordinate system which origin at the 
center of the LS sphere.
\begin{equation}
\label{eq:track}
\vec{track}=(x_{\scriptsize\textrm{in}}, y_{\scriptsize\textrm{in}}, z_{\scriptsize\textrm{in}}, p_{x}, p_{y}, p_{z}).
\end{equation}

Muon reconstruction is the estimation of the 6 independent track parameters $\vec{track}^{\scriptsize\textrm{rec}}$
from PMT measurements of the numbers of photoelectrons ($\vec{Q}$), fastest light arrival times ($\vec{T}$)
as well as the positions of all the PMTs ($\vec{R}$).	
\begin{equation}
\label{eq:func}
\vec{track}^{\scriptsize\textrm{rec}}=f(\vec{Q}, \vec{T}, \vec{R}). 
\end{equation}

Reconstruction performance is assessed using the 
angle $\alpha$ between the true track and the reconstructed track 
and the difference between the distance to CD center $\Delta$D between 
the true and reconstructed track.

\subsection*{Muon tracking and selection with Top-Tracker}

Comparisons of TT reconstructed muon tracks with MC truth tracks 
show that it is necessary to apply track quality criteria 
to select well reconstructed muons. Table~\ref{tab:cut} shows 
the reconstruction performance for a series of selection 
criteria on the number of the hit TT layers ($N_{\scriptsize\textrm{point}}$), 
the least squares value of TT reconstruction ($\chi^2$) and the number of
activated scintillator bars in the X, Y direction of the layer ($N_{\scriptsize\textrm{channel}}$).

The reconstruction performance is substantially improved within the 
high quality selection. Further quality criteria tested
did not provide significant performance gains. The highest quality selection 4    
was used for the network training with TT tracks.


\begin{table}[h]
\centering
\caption{TT reconstruction performance for a sequence of selection criteria.}
\label{tab:cut}
\begin{tabular}{l|llll}
\hline\noalign{\smallskip}
Selection      & 1     & 2     & 3       & 4 \\\hline
Events         & 145K  & 130K  & 46K     & 40K  \\
$N_{\scriptsize\textrm{point}}$ Cuts   & -     & $>$2  & $>$2    & $>$2 \\
$\chi^2$ Cuts      & -     & -     & $<$0.2  & $<$0.2  \\
$N_{\scriptsize\textrm{channel}}$ Cuts & -     & -     &  -      & $<$8 \\
$\alpha$ Mean / degree & 0.30  & 0.23  & 0.21  & 0.20  \\
$\alpha$ RMS / degree  & 1.28  & 0.58  & 0.30  & 0.16  \\
$\Delta$D Mean / mm    &  1.1  &  0.8  &  0.6  &  0.3  \\
$\Delta$D RMS / mm     &  321  &  157  &   91  &   75  \\
\noalign{\smallskip}\hline
\end{tabular}
\end{table}

\subsection*{Rotation method based data augmentation}

\begin{figure}[h]
		\centering
		\subfigure[Injection points distribution of TT reconstruction]{
			\includegraphics[width=1\linewidth]{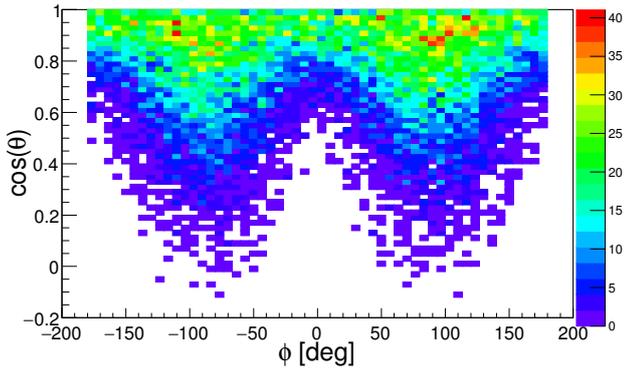} }
		\subfigure[Injection points distribution of MC truth tracks]{
			\includegraphics[width=1\linewidth]{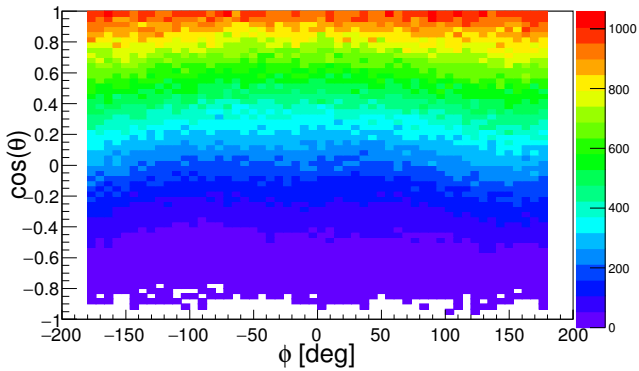} }  
		\caption{The two dimension distribution of ($\cos\theta$, $\phi$), where ($\theta$, $\phi$) are the injection positions at the surface of the LS sphere in spherical coordinates. The injection positions are at the north pole of LS sphere when cos($\theta$) = 1 ($\theta$ = 0 deg). The distribution modulated with and without TT geometry and reconstruction is shown in (a) and (b) respectively. In Fig (a), the non-uniform distribution along $\phi$  is influenced by the TT geometry. In Fig (b), the distribution is not uniform along $\phi$, because the underground muon distribution was modulated by the mountain profile above.}  
		\label{fig:inj}   
	\end{figure}

As the TT partially covers the water pool it is only able to 
reconstruct muons over a limited solid angle region.  
Figure~\ref{fig:inj} shows the distributions of 
CD sphere injection points for reconstructed TT muons in the upper plot 
and for the the MC truth tracks in the lower plot.
A rotation method has been developed that effectively extends the angular range 
of the training sample using a rotational symmetry assumption.

\begin{figure}[htb]
		\centering
		\includegraphics[width=0.8\linewidth]{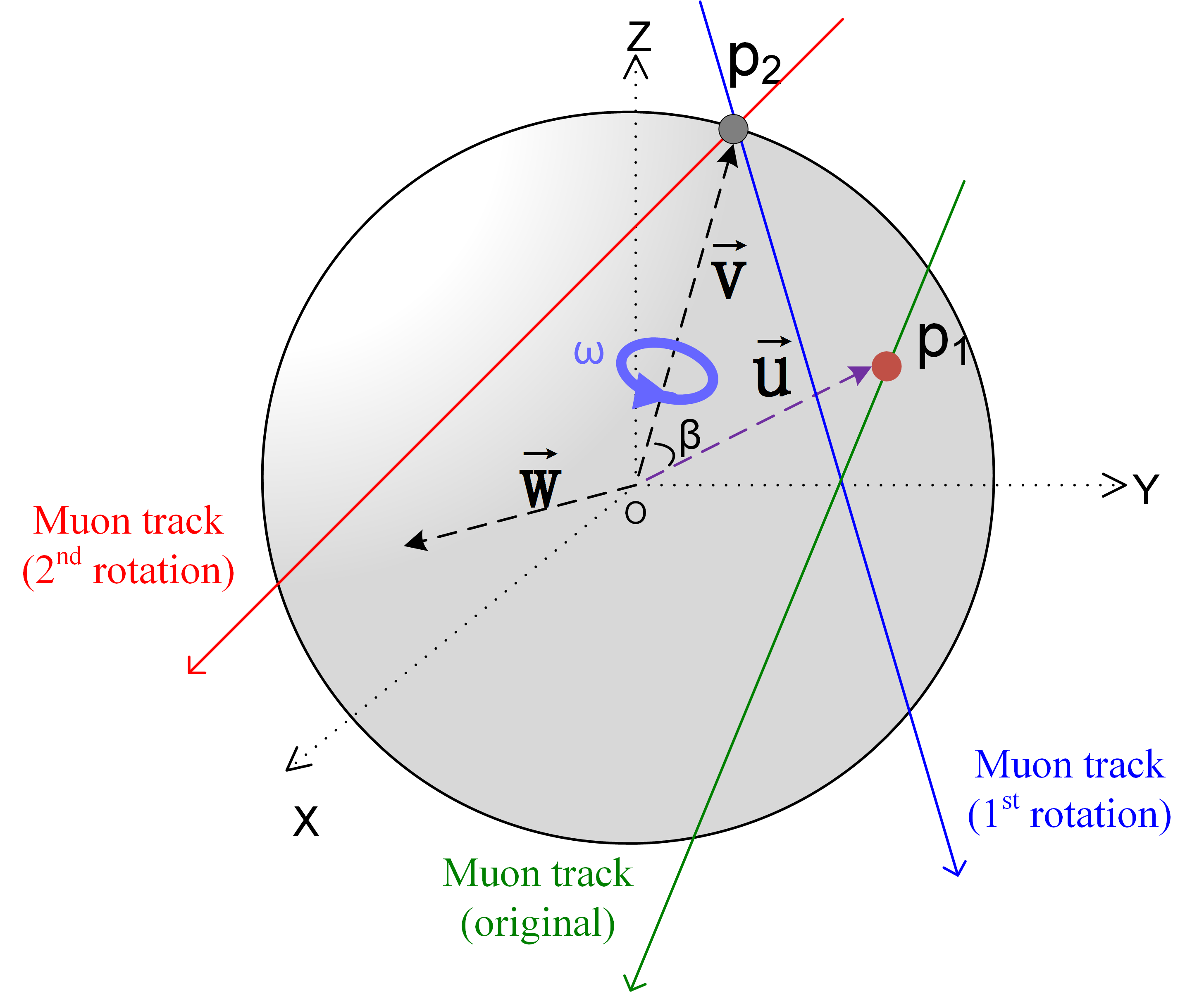}
		\caption{
Schematic diagram of the rotation method. The original muon is shown in green.
First the track is rotated along the $\vec{w}$ with angle $\beta$, where
$\vec{w}=\vec{u}\times\vec{v}$. After the first rotation, the track in blue is
then rotated around $\vec{v}$ with angle $\omega$. The final track is shown in
red. The detector is also rotated using the same rotation matrix.  }
		\label{fig:r1}
	\end{figure}

The rotation method is illustrated in Fig.~\ref{fig:r1} showing 
how an event is rotated and a new event is generated based on the reconstructed tracks. 
A randomized rotation matrix of each event is defined considering a new injection point 
and rotation angle $\omega$ of the track around the vector $\vec{v}$.
The track and the positions of PMTs are rotated together using the same rotation matrix.
After the randomized rotation, a rejection sampling is applied to get a new position that conforms to the expectation of the distribution.

\begin{figure}[h]
		\centering
		\includegraphics[width=1\linewidth]{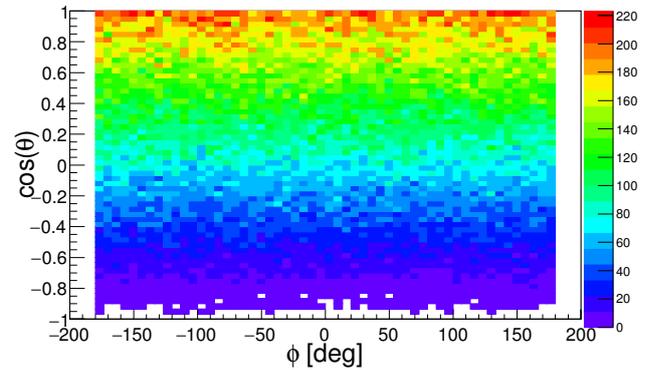}
		\caption{Injection point distribution after rotation}
		\label{fig:rot-data}
	\end{figure}

For muon events
reconstructed by TT, after using the rotation method, the distribution of the 
injection point is shown in Fig.~\ref{fig:rot-data}.
Comparing the Fig.~\ref{fig:inj} (b) and Fig.~\ref{fig:rot-data}, it can be
found that the injection point distribution after rotation is similar to the
one from cosmic muon in CD which basically extends the ranges of injection points.

\subsection*{Dataset Composition}

Application of CNN models to muon reconstruction requires the fixed PMT positions
and measurements of PMT charge and times to be converted into image arrays.
All PMT positions (x,y,z) are converted into two-dimensional spherical coordinates
($\theta$,$\phi$) using the equidistant cylindrical projection.
%
%
	
\begin{figure}[h]
		\centering
		\subfigure[ Image of channel Q]{
			\includegraphics[width=1\linewidth]{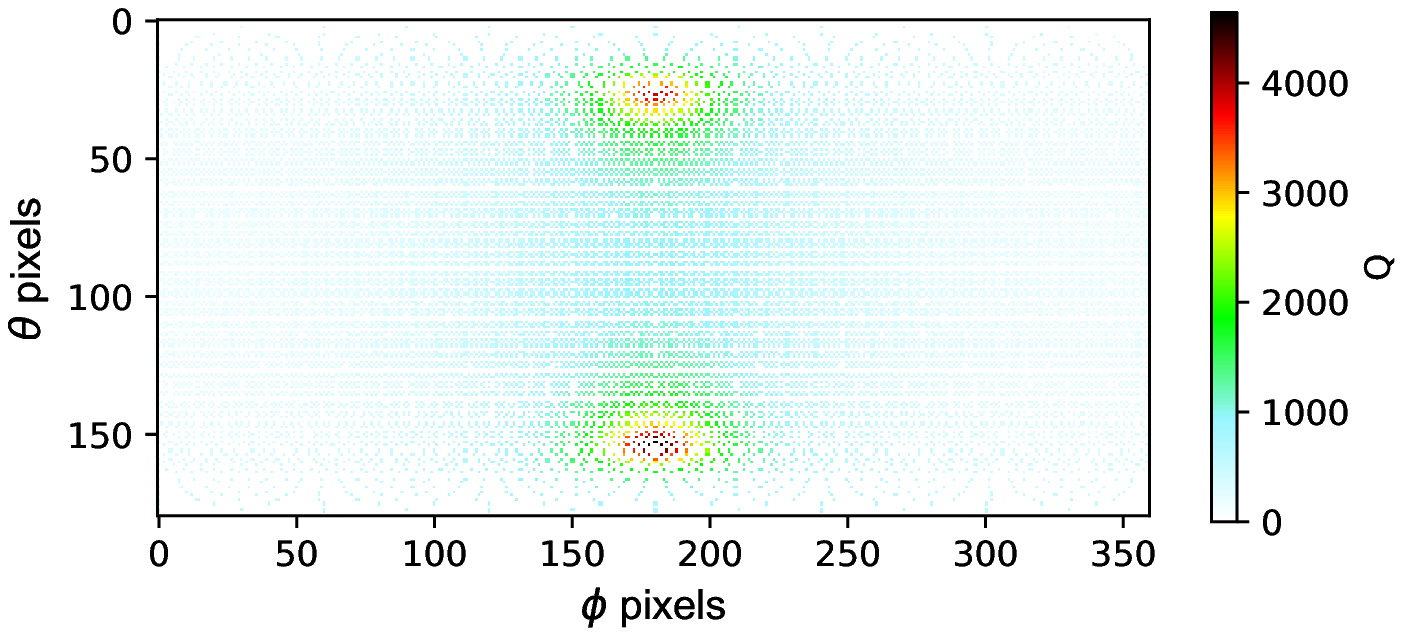} }
		\subfigure[  Image of channel T]{
			\includegraphics[width=1\linewidth]{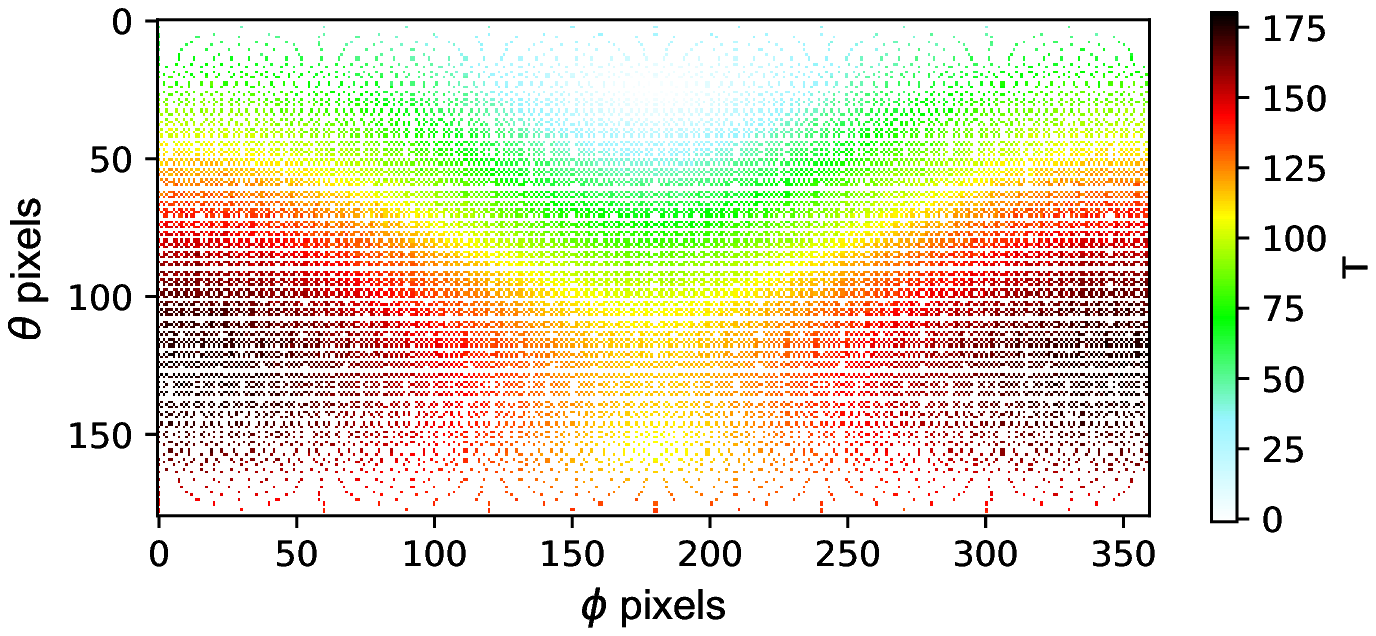} }
		\caption{%
Charge and time channel images for a vertical downwards muon track with entry point at spherical coordinates (25,180) degrees
on the CD sphere. The color scale from blue through to red represents increasing values. %
}
		\label{fig:pos}   
	\end{figure}

Example images of the charge Q and time T channels at the PMT pixel positions 
are shown in the Fig.~\ref{fig:pos}. 
The charge channel has two clusters corresponding to the muon entry
and exit points at the CD sphere surface and the time T channel has 
a valley of low values in the region of the muon entry point.
These two channels are combined together in an image pixel array, 
with each pixel having two channel values with the values of charge Q and time T,
while the values for pixels that do not correspond to PMTs are set to zero.
Then the feature standardization \cite{pal2016preprocessing} is applied to each channel of the image pixel array 
for the normalization of the image before entering the neural network:
\begin{equation}
X' = (X - \mu) / \sigma
\end{equation}
where $X$ is the original data, $X'$ is the normalized data, 
$\mu$ is the mean of $X$ and $\sigma$ is the standard deviation of $X$.

In the training, the label of the direction vector ($p_{x},p_{y},p_{z}$) and intersection position vector ($x_{\scriptsize\textrm{in}}, y_{\scriptsize\textrm{in}}, z_{\scriptsize\textrm{in}}$) have been normalized to unit vectors, respectively.

The training and testing samples used in this study are shown in Table~\ref{tab:dataset}. 
The sample of directly simulated muons is split into training, validation
and testing samples named ``Training MC'', ``Validation MC'' and ``Testing MC'' in the table.
The training data set obtained by TT reconstruction and rotation is called ``training TT$\_$rot''. 

	\begin{table}[h]
		\centering
		\caption{Dataset}
		\label{tab:dataset}
		
		\begin{tabular}{llll}
			\hline\noalign{\smallskip}
			Dataset & Events & Image size & Label\\
			\noalign{\smallskip}\hline\noalign{\smallskip}
			Training MC    &300K&180x360 QT&MC truth\\
			Training TT$\_$rot&300K&180x360 QT&TT reconstruction\\
			Validation MC     &20K &180x360 QT&MC truth\\
			Testing MC     &20K &180x360 QT&MC truth\\
			\noalign{\smallskip}\hline
		\end{tabular}
		
	\end{table}
	\section*{Network architecture}
The use of artificial neural networks to solve complex problems has been
explored since the 1940s. In recent years, with the dramatic increase in
available computing power, it has become feasible to use computationally
intensive neural networks with many inner layers. These neural networks with
many computational layers are called deep neural networks (DNN), which can
learn various potential features in large amounts of data. Convolutional neural
network (CNN or ConvNet) are one form of deep neural network. CNN consist of one or more
convolutional layers and a top fully connected layer, as well as associated
weights and pooling layers. This structure makes convolutional neural networks
suitable for processing two-dimensional image data and extracting features from the data
efficiently.

\subsection*{Model architecture}
\begin{figure*}
		\centering
		\includegraphics[width=0.8\linewidth]{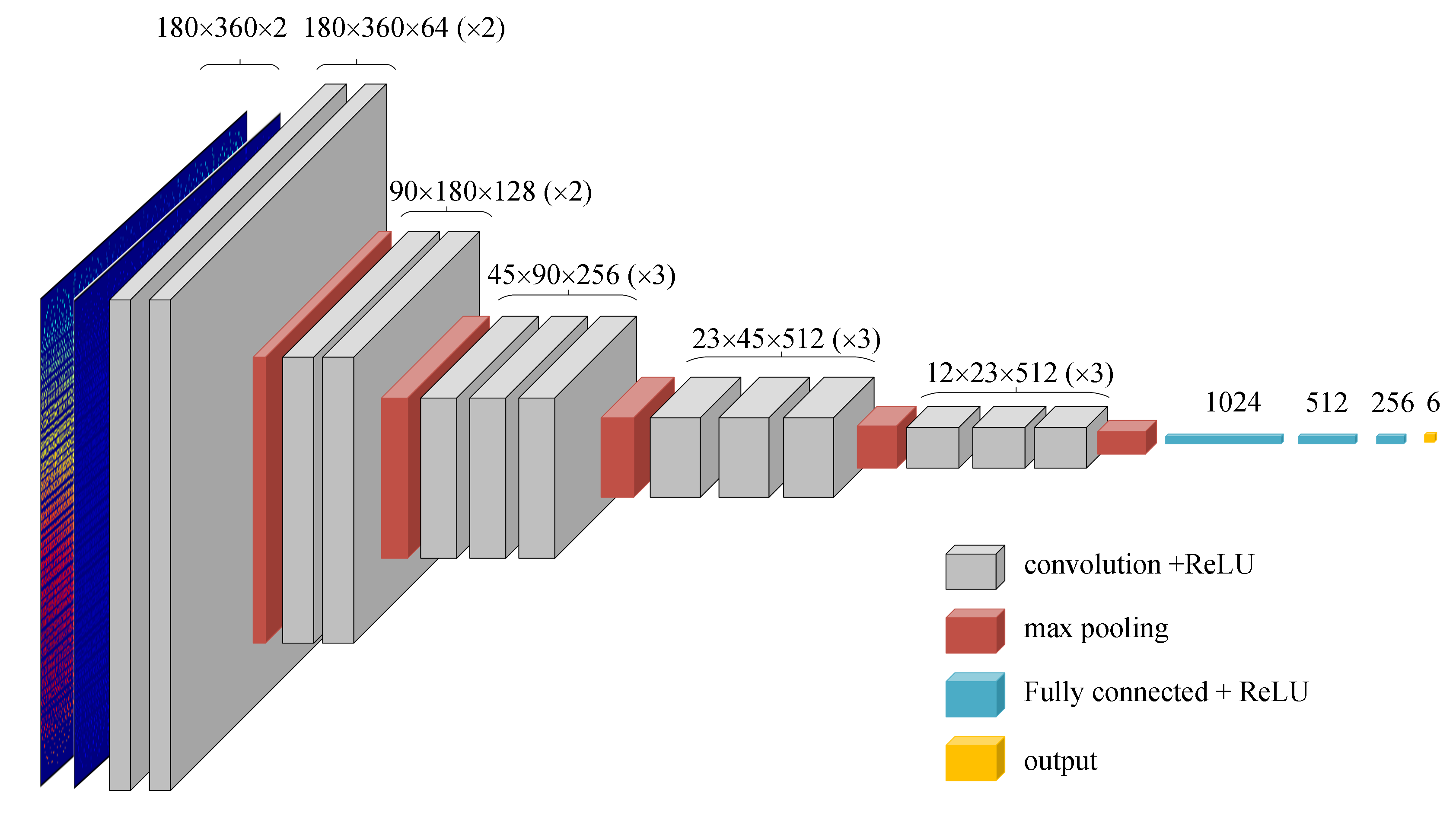}
		\caption{Convolutional neural networks configuration adapted from VGGNet16, which is widely used to extract image features and image classification.}
		\label{fig:conf}
	\end{figure*}

%
The network model architecture adopted in this study and illustrated in Fig.~\ref{fig:conf}
is based on the VGGNet16 \cite{simonyan2014very} configuration.
It is composed of an input layer, convolutional layers, activation layers, pooling
layers and fully connected layers. The input layer is the 2-channel (Q and T) image 
converted from the muon event data.

The principal elements in this model are blocks of three convolution layers
with a 3$\times$3 convolutional kernel followed by a pooling layer (max pooling)
\cite{scherer2010evaluation}. The convolutional layers are used to perform data feature
extraction. The max pooling layer is used for signal downsampling, reducing the
data size input to the next layer and also reducing the impact of local data
on the result which makes the results more robust. As shown in the study~\cite{simonyan2014very} 
the use of three 3$\times$3 convolutional layers has several advantages.
Firstly this architecture has the same large effective receptive field as a 7$\times$7
convolutional layer, but decreases the number of learning parameters by almost a factor of two.
Secondly, it can inject more non-linearity operations between the two
convolution layers which enhanced the ability to learn features.
	
The fully connected hidden layers convert the abstract feature signal into the output of the prediction task. After some tries, the number of neurons at each fully connected hidden layer is chose as following: 1024, 512, 256. Comparing with original VGGNet, this kind of setting not only gives better muon reconstruction performance but also reduces the number of parameters in fully connected hidden layers by 75\%. 



Finally, a fully connected output layer with 6 nodes which corresponding to the muon track parameters is applied to obtain the predictions of the network. 
The six output values of the network are the predicted values of track
entry position into CD ($x_{\scriptsize\textrm{in}}$, $y_{\scriptsize\textrm{in}}$, $z_{\scriptsize\textrm{in}}$) and track direction ($p_{x}, p_{y}, p_{z}$) of the muon. 

\subsection*{Loss function}

The loss function is an indicator used to measure the difference between the model prediction and the true values
which is used for the optimization of the model. The loss function used in this study, shown in Eq. \ref{eq:qtr}, 
consists of two terms. The first term represents the error between the true label and predicted label of the muon track. 
The second term provides a regularization penalty \cite{phaisangittisagul2016analysis} which acts to avoid overfitting of the model and contains
$\lambda$ coefficient scale and a summation over the weights of the model $W_{j}$.
	
	\begin{equation}
	\label{eq:qtr}
	Loss =\sum_{i}(||\vec{track}^{\scriptsize\textrm{predict}}_{i}-\vec{track}^{\scriptsize\textrm{truth}}_{i}||^2)+\sum_{j} \lambda ||W_{j}||^2
	\end{equation}

The training process acts to reduce the loss function resulting in the predicted values 
approaching closer to the actual values. With muon reconstruction this corresponds to 
improving the spatial and angular resolution of the reconstruction.

\subsection*{Implementation and training}
The network used in this article is implemented using the TensorFlow~\cite{Abadi:2016kic} 
open source framework and its Python interface. The hardware environment uses
the NVIDIA Tesla V100 GPU. Network training uses 2 GPUs for parallel
computing. Using Mini-Batch \cite{orr2003neural} training strategy, the size of batch
is 256 (256 samples are trained at the same time in each iteration), which is
divided into two smaller batches and computed in parallel on 2 GPUs. Two GPUs
provides a speedup of 1.7 times, as compared to using a single GPU. On a system
equipped with two NVIDIA Tesla V100 GPUs, training a single network took about 70
hours depending on the architecture.
	
The training neural network uses the stochastic gradient descent method to
calculate the gradient of the loss function in the weight parameter space, and
then advances the weight parameters by a certain step in the direction of the
negative gradient. After reaching the set number of iterations or the set error
accuracy, the weight parameter is considered to have reached the optimum position where
the loss function is the smallest. The step size in the gradient
descent process is the learning rate.
After some studies, the following optimized learning rate schedule is used: the learning rate is initialized to be 0.1 at the beginning of training and is decreased by a factor of 10 after each 50,000 training steps. 

The loss values from training dataset and validation dataset are monitored during the network training. It is found that after 200,000 training steps, the loss values from training dataset and validation dataset are almost stable which means the network has reached the optimized status. Besides, during the training the loss from validation dataset decreases smoothly which indicates the training process does not  suffer from overfitting (a phenomenon that a model has over-learned features specific to the training dataset that are not present in general datasets, which is harmful to the model).

\section*{Performance study}
	
\subsection*{Reconstruction resolution and efficiency}
The network proposed in this paper is separately trained on dataset ``Training
TT$\_$rot'' and ``Training MC'' which is used for verifying the rotation method.
After training, they are tested by dataset ``Testing MC'' for performance
analysis. 
After trying some fitting functions for the distribution of $\alpha$, it is found that the fitting effect is good using the improved Landau function~\cite{Cao:2010zzd}
$f(x) = p_0 \exp(-p_3(\lambda+\exp(-\lambda)))$, 
where $\lambda = \frac{x - p_1}{p_2}$.
The full width at half maximum (FWHM) of the fitted distribution is defined as the resolution of $\alpha$. 
The distribution of reconstructed $\Delta$D is fitted to a Gaussian function and its resolution is obtained from the sigma of the Gaussian.

\begin{figure}[h]
		\centering
		\subfigure[$\alpha$ distribution fitted by improved Landau function]{
			\includegraphics[width=1\linewidth]{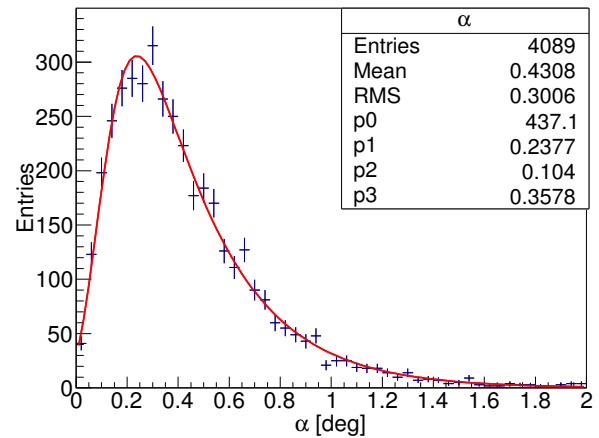} }
		\subfigure[$\Delta$D distribution fitted by Gaussian function]{
			\includegraphics[width=1\linewidth]{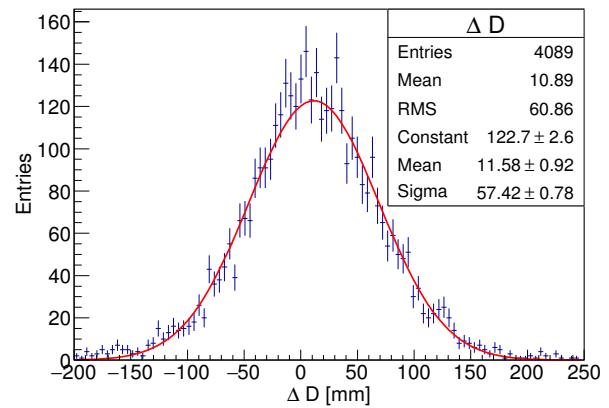} }  
		\caption{$\alpha$ and $\Delta$D reconstructed from the events where the track is 14.7 m to 16.7 m from the center of CD}  
		\label{fig:per2}   
	\end{figure}

Figure \ref{fig:per2} (a) and (b) show the distributions and fitting results of
$\alpha$ and $\Delta$D obtained by using ``Training TT$\_$rot'' as training data
when the distance of muon track from the center of CD ($D_{mc}$) is 14.7~m-16.7~m. 
Good fits have been achieved for both the angular and spatial resolutions. 

\begin{figure}[htbp]
		\centering
		\subfigure[Most probable value(MPV) and full width at half maxima(FWHM) value of $\alpha$ versus $D_{mc}$]{
			\includegraphics[width=1\linewidth]{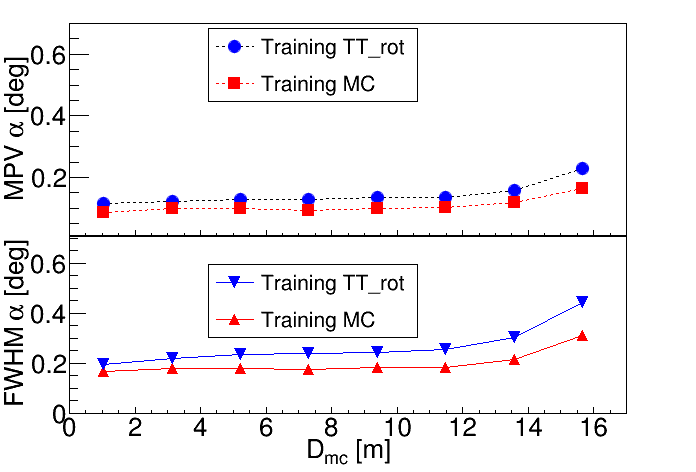} }
		\subfigure[Mean and sigma value of $\Delta$D versus  $D_{mc}$]{
			\includegraphics[width=1\linewidth]{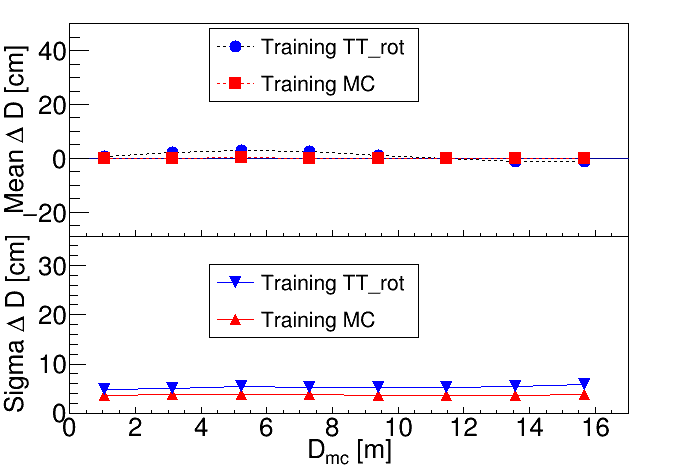} }  
		\caption{Performances of reconstruction with various training data.}  
		\label{fig:per1}   
	\end{figure}

Figure \ref{fig:per1} shows how the reconstructed $\alpha$ and $\Delta$D vary with
$D_{mc}$. Comparing the performance obtained when training with the  ``Training TT$\_$rot'' and
``Training MC'' datasets we observe:
\begin{itemize}
		\item %
For the ``Training TT$\_$rot'' dataset the angular bias of $\alpha$ is less
than 0.4 degrees, and the spatial bias of $\Delta$D is less than 3 cm. Also the results 
from the ``Training TT$\_$rot'' and ``Training MC'' are very similar suggesting that using TT reconstructed tracks 
with experiment data will provide a realistic way to train the network that does not depend on the simulation.
		\item %
The sigma (or resolution) of $\alpha$ and $\Delta$D increases with the
$D_{mc}$. When the muon track is close to the edge of CD, the position of the
incident point and the exit point are very close to each other which makes an accurate reconstruction 
of the track direction difficult. This degradation of angular reconstruction at large distances 
from the CD center is also seen with the existing muon reconstruction methods\cite{Genster:2018caz,Zhang:2018kag}.
Overall, the angular resolution is better than 0.6 degrees and the spatial resolution is better than 10 cm.
\end{itemize}

\begin{figure}[ht]
		\centering  
		\includegraphics[width=1.0\linewidth]{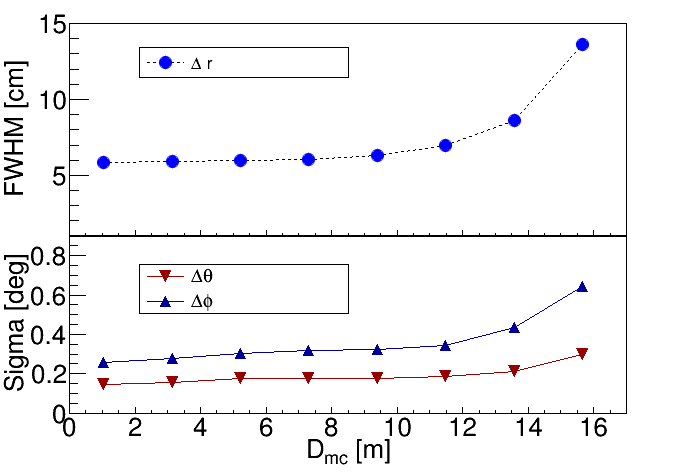} 
		\caption{Reconstruction spatial resolution and angular resolution}
		\label{fig:per3}   
	\end{figure}

In order to understand our muon reconstruction method better, figure~\ref{fig:per3} provides another check of the muon reconstruction performance obtained with the dataset ``Training TT$\_$rot''. 
Here the $\Delta r$ resolution is expressed as FWHM of $\Delta r=|\vec{r}^{\scriptsize\textrm{predict}}_{\scriptsize\textrm{in}} - \vec{r}^{\scriptsize\textrm{truth}}_{\scriptsize\textrm{in}}|$ which means the difference between the true and predicted positions of the muon injection point at LS surface.
The direction resolution is showed by the sigma of the distribution of $\Delta\theta$ and $\Delta\phi$ which express the differences of $\theta$ angles and $\phi$ angle between the reconstructed and true direction of muon track, respectively.
Form figure~\ref{fig:per3}, one can see the results are normal and there are no anomalies.

\begin{figure}[ht]
		\centering  
		\includegraphics[width=1.0\linewidth]{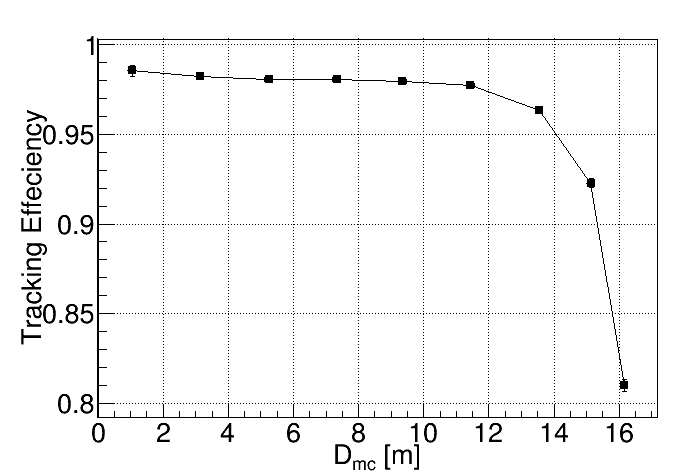} 
		\caption{Reconstruction efficiency for various muon track distances to the CD center.}
		\label{fig:p3}   
	\end{figure}

The track reconstruction efficiency is defined as the ratio of 
well reconstructed tracks to all tracks, with tracks  
regarded as well reconstructed when the deviations of all parameters 
are less than five times their standard deviations. 
Figure \ref{fig:p3} presents the reconstruction efficiency 
for a range of track distances from the CD center. For distances 
of less than 14~m the tracking efficiency is greater than 95\%, dropping 
to 86\% for tracks at the edge of the CD.
%
\subsection*{Reconstruction time cost}
The time cost of muon reconstruction using three different approaches 
is presented in Table~\ref{tab:time}. The time for the CPU implementation 
of the neural network method presented in this paper is found to 
be improved by more than a factor of two 
compared to the fastest light method.
The GPU implementation of the neural network method provides a huge processing speed 
improvement of a factor of one hundred compared to the CPU implementation.  
	\begin{table}
		\centering
		\caption{Reconstruction speed.}
		\label{tab:time}
		\begin{tabular}{cccccc}
			\hline\noalign{\smallskip}
			\makecell[c]{Method }&\makecell[c]{ Hardware }& \makecell[c]{Time per event [ms]}\cr
			\noalign{\smallskip}\hline\noalign{\smallskip}
			Fastest Light    &\makecell[c]{ CPU\\E5-2650 v4} & 5000\cr
			\makecell[c]{Neutral Network \\ (batch size=1)} &\makecell[c]{CPU\\E5-2650 v4} & 2000\cr
			\makecell[c]{Neutral Network \\ (batch size=1)} &\makecell[c]{GPU\\Tesla V100} & 20\cr
			\noalign{\smallskip}\hline
		\end{tabular}
	\end{table}
\section*{Discussion}

The muon reconstruction accuracy of the deep learning model
largely relies on the amount and the diversity of data available
during training. Applying rotation to injected muon tracks can
be effectively used to augment the training data. However the
disadvantage with the rotation method is that it could lead to
small changes in geometry that can cause a muon track
represented by an image to lose part of its original features,
which might possibly degrade the performance of reconstruction.
Studies show that the performance discrepancy between the
rotation method and the MC truth method is at the level of a few
percent, which is still in an acceptable range. In order to further improve the performance of the CNN model, more sophisticated data augmentation method, such as data augmentation using GANs \cite{tanaka2019data}, can be investigated and potentially used to generate artificial training data in the future.

The muon reconstruction results presented in the previous
section are obtained in an ideal scenario. With the real
experimental data, there are many factors that could possibly
degrade the performance of the model. The possible influence
factors are as follows: the dark noise created by the PMT,
electronics saturation, the deformation of acrylic sphere etc.

When the experiment starts data taking, it takes time to
complete the tuning of Monte Carlo simulation to make the
simulation consistent with the data. During this period, deep
learning model can be obtained by training with the augmented
data sample. As soon as the reliable simulation data are
available, the training can move to use the full simulation
data.

\section*{Conclusion}

In this study a deep learning convolutional neural network approach is applied to muon track reconstruction in the JUNO central detector.
Achieving good agreement between the real JUNO experimental data and the simulated MC data is expected to require extensive efforts to develop a detailed understanding of many small effects corresponding to improvements to the models of the scintillation, re-emission, photon propagation into PMTs and the electronics. 
Reconstruction techniques which are not dependent on the availability of finely tuned simulation data are highly beneficial. In this study the track information reconstructed by the top tracker is used for network training instead of relying on Monte Carlo truth information.
Due to the limited angular coverage of the top tracker a randomized event rotation technique has been developed which succeeds to generate sufficient training data by making the assumption of rotational symmetry. 
This dataset was used to train a convolutional neural network with promising reconstruction performance. The reconstruction efficiency is higher than 95\% with the distance less than 14 meters of the track to CD center. The angular resolution is better than 0.6 degrees and the spatial resolution is better than 10 cm, and these results demonstrate that muon reconstruction a convolutional neural network is able to provide an performance that meets the requirements. Compared to existing methods\cite{Zhang:2018kag}, the CNN model does not need to introduce additional real data correction.
In addition, the GPU implementation of the technique has the advantage of a processing speed more than 100 times faster than conventional techniques.

%
%
%
	
%
%
\end{document}